\documentclass[12pt]{article}
\usepackage{epsf}

\newcommand{\bee}{\begin{equation}}
\newcommand{\ee}{\end{equation}}
\newcommand{\bea}{\begin{eqnarray}}
\newcommand{\eea}{\end{eqnarray}}
\newcommand{\R}{\rm I\kern-.2emR}
\newcommand{\C}{\rm \kern.25em\vrule height1.4ex
depth-.12ex width.06em\kern-.31em C}
\newcommand{\N}{{\rm I\kern-.16em N}}
\newcommand{\Z}{{\rm Z\kern-.35em Z}}
                                                      
\begin{document}                                                                
\begin{flushright}
MPI-PhT/2001-43\\
\end{flushright}
\bigskip\bigskip
\begin{center}
\Huge{Testing Asymptotic Scaling and Nonabelian Symmetry Enhancement.
}
\end{center}
\centerline{\it Adrian Patrascioiu}
\centerline{\it Physics Department, University of Arizona}
\centerline{\it Tucson, AZ 85721, U.S.A.}
\centerline{\it e-mail: patrasci@physics.arizona.edu}
\vskip5mm
\centerline{\it Erhard Seiler}
\centerline{\it Max-Planck-Institut f\"ur Physik}
\centerline{\it (Werner-Heisenberg-Institut)}
\centerline{\it F\"ohringer Ring 6, 80805 Munich, Germany}
\centerline{\it e-mail: ehs@mppmu.mpg.de}
\bigskip \nopagebreak

\begin{abstract}
We determine some points on the finite size scaling curve for the 
correlation length in the two dimensional $O(3)$ and icosahedron spin 
models. The Monte Carlo data are consistent with the two models 
possessing the same continuum limit. The data also suggest that the 
continuum scaling curve lies above the estimate of Kim \cite{kim} and 
Caracciolo et al \cite{car} and thus leads to larger thermodynamic 
values of of the correlation length than previously reported.
\end{abstract}
\vskip2mm
In 1993 Kim \cite{kim} proposed using finite size scaling to obtain 
the thermodynamic value of the correlation length $\xi$ in the
two dimensional ($2D$) nonlinear $\sigma$ model at given inverse 
temperature $\beta$. He claimed that he could predict correctly (less 
than 2\% error) values as large as 15000 from measurements taken on 
lattices not larger than $283\times 283$. Kim's method was adopted by
Caracciolo et al \cite{car}, who, after refining it and collecting data at 
a 180 pairs of $L$ and $\beta$ values claimed to have such control over the
scaling curve that they could predict values of the correlation length
as large as $10^5$ even though their largest lattice was only 
$512\times512$.

Of course at fixed $x=\xi(L)/L$ the value of $\xi(2L)/\xi(L)$ depends upon
$L$ and the continuum value is the limit of this ratio as $L\to\infty$.
Both Kim and Caracciolo et al claimed that within their statistical 
accuracy, they could not detect any systematic drift of $\xi(2L)/\xi(L)$
with $L$ within the range of values of $L$ they studied; so they concluded
that they had reached the continuum limit. 
(The latter authors also investigated in more detail the approach to the
continuum limit in a different model, the two-dimensional $SU(3)$
principal chiral model \cite{sokal} and convinced themselves that their
continuum extrapolation was supported by their data (private communication
from Alan Sokal)). 

Shortly after the appearance of these papers,
we criticized both Kim's paper \cite{comk} and the Caracciolo et al paper
\cite{coms} by pointing out that we know rigorously \cite{jsp} that in the true 
continuum limit large spin fluctuations have to occur for any value
of $x>0$. Indeed it can be shown that the spins must become decorrelated
over distances that are fixed, nonvanishing multiples of the infinite
volume correlation length, while in the regime in which these authors took
their data, at least at $x>0.75$, the lattices were so well ordered that 
second order lattice perturbation theory (PT) could reproduce their 
Monte Carlo results within the errors.

In this paper we significantly improve Caracciolo et al's statistics,
and go to larger lattices (their largest $L$-$2L$ pair has $L=256$, ours
$640$). Thereby we find (for $\xi(L)/L=0.5$)  some statistically 
significant corrections to scaling even at the largest $L$ values, which
in our view signifies that the data of Caracciolo et al were rather
far from the continuum limit even at $x=0.5$ (of course, it is to be
expected, and the present study corroborates that, that the continuum
limit for larger values of $x$ is reached at larger values of $L$).

Our interest in this problem was revived by the recent discussions 
of symmetry enhancement of discrete nonabelian groups.
Namely both we \cite{p} \cite{ps} and Hasenfratz and Niedermayer
\cite{hn} pointed out that numerics suggest that the dodecahedron and
icosahedron spin models have the same continuum limit as the $O(3)$
model. Hasenfratz and Niedermayer interpreted this as meaning that these
discrete spin models were asymptoticaly free (AF).
In fact in our paper we \cite{ps} had measured the L\"uscher-Weisz-Wolff
running coupling constant \cite{lww} and showed that it did not vanish 
at the critical point, as required by AF. A theoretical argument
was also published shortly thereafter \cite{ital} claiming to show
the impossibility of the Hasefratz-Niedermayer scenario of AF in
a discrete spin model.

This latter paper contains an interesting suggestion: the observed
agreement between the discrete spin models and $O(3)$ is a transient
phenomenon, which should disappear at larger correlation length. The 
observation stems from the fact that if one accepts the standard 
scenario of $O(3)$ being AF, then the discrete spin models could be 
regarded as a perturbed $O(3)$ model, and in the accepted PT scheme 
around the Gaussian fixed point, this would be a {\it relevant}
perturbation, hence the two models could not possibly be equivalent.
However the authors find that this perturbation becomes relevant
only for $\beta$ sufficiently large. They translate their estimate of 
this value of $\beta$ into a correlation length of approximately 200 
and suggest that perhaps the numerics will show that as the correlation
length is increased, for continuum observables, the agreement between
$O(3)$ and the icosahedron or dodecahedron spin models improves for a
while, then, as $\xi$ exceeds approximately 200, it starts deteriorating.

We decided to investigate numerically this possibility by comparing the
finite size scaling curve of $\xi(L)$ in the two models. We studied the
icosahedron spin model and the $O(3)$ model at $x=0.25$, $x=0.5$ and
$x=0.75$. For the icosahedron we also took data at our estimated
$\beta_{crt}=1.803$ \cite{ps}. The results are recorded in Tabs.1-7 and
displayed in the figures. 

Before discussing the figures, let us specify in detail what we did and 
summarize what we find. Our systems consist of two dimensional 
square arrays of spins of size $L\times L$ with periodic boundary 
conditions. The spin at each site is of unit length and takes values
either on the sphere $S^2$ or on the vertices of an inscribed regular
icosahedron. Each spin interacts ferromagnetically only with
its 4 nearest neighbours at inverse temperature $\beta$. 

We used the same definition of the correlation length $\xi$ as 
\cite{car}: let $P=(p,0)$, $p={2n\pi\over L}$, $n=0,1,2,...,L-1$. Then
\bee
\xi={1\over 2\sin(\pi/L)}\sqrt{(G(0)/G(1)-1)}
\ee
where
\bee
G(p)={1\over L^2}\langle |\hat s(P)|^2\rangle;\ \
\hat s(P)=\sum_x e^{iPx} s(x)
\ee

At a given $L$, we adjusted $\beta$ so that the ratio $\xi(L)/L$ was
approximately 0.25, 0.5 respectively 0.75 (we allowed differences only smaller
than $2\times 10^{-3}$). Leaving $\beta$ unchanged, we then doubled L and
measured $\xi(2L)$ and therefore $\xi(2L)/\xi(L)$ at our $x$. 
(We used the slope of the scaling curve of \cite{car} to correct for the
fact that our $\xi(L)/L$ is not exactly equal to the desired values
of 0.25, 0,.5 and 0.75 ; since this correction is tiny, it does not
matter whether their scaling curve represents exactly the truth).
In our study $L$ was varied from 20 to 640 or to 1280 and the data were
produced using the usual one cluster algorithm. For each $\beta$ $L$ pair
we performed several runs (up to 200), one run consisting of 100,000
clusters used for thermalization and 1,000,000 for taking
measurements. Each run started from a new randomly chosen configuration.
The errors were computed from the results produced by the different runs,
using the jack-knife method.

In Fig.\ref{step5} we show our results for $O(3)$ (full squares) and 
for the icosahedron (open circles) at $x=0.5$. For comparison, we also 
plotted the results of Caracciolo et al for $O(3)$ (full triangles). 
The choice of the abscissa $1/(\ln L +c)$, was motivated by the fact 
that in studying lattice artefacts for the LWW step scaling function 
we found that such an ansatz seemed to describe the data pretty well 
(see ref.\cite{lat} Fig.7). The value of the parameter $c=0.7$ was 
obtained from a joint fit to our $O(3)$ and icosahedron data with a 
common limiting value for $L\to\infty$; it is of acceptable quality 
($\chi^2/dof=11.6/8$). The solid curve is a Symanzik type fit 
($a+b/L^2+c\ln L/L^2$) to the $O(3)$ data with, however, an unacceptable
$\chi^2/dof=10.7/3$. Several facts are suggested by this figure:

\begin{figure}[htb]
\centerline{\epsfxsize=10.0cm\epsfbox{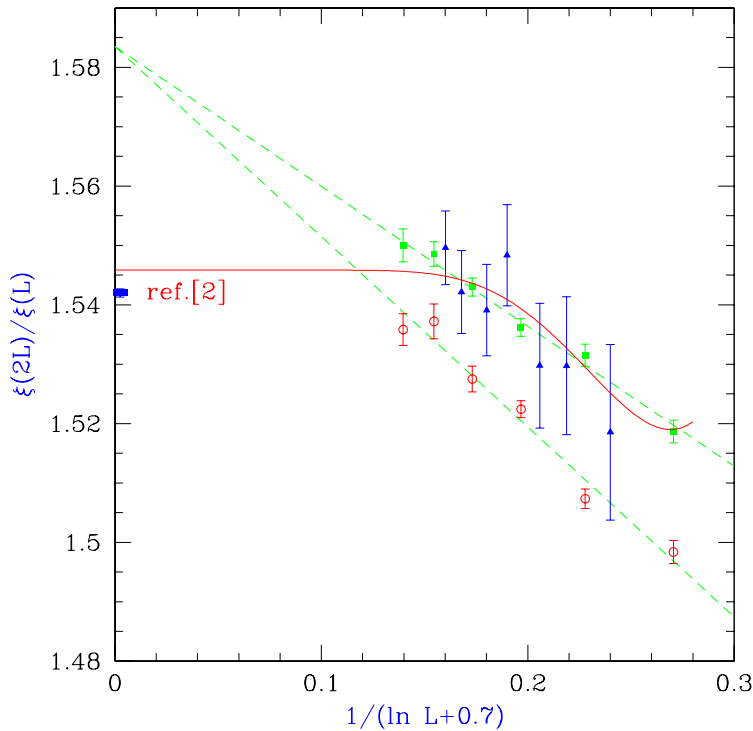}}
\caption{Scaling function $\xi(2L)/\xi(L)$ versus
$1/(\ln L+0.7)$ for $O(3)$ (full triangles: ref.\cite{car} and full
squares  : our data) and
icosahedron (open circles) at $\xi(L)/L=0.5$. The straight lines are
fits.}
\label{step5}
\end{figure}

\begin{itemize}
\item Our MC data agree very well with those of Caracciolo et al. However
while the largest $L$ value investigated by these authors was $L=256$, ours
is $L=640$. Also our error bars are much smaller.
\item The $O(3)$ data suggest a systematic increase of $\xi(2L)/\xi(L)$
with $L$. This is true about both our data and those of Caracciolo et al,
the latter however, having much larger error bars, could also be interpreted
as showing no $L$ dependance for $L>64$.
\item The data suggest that the continuum value is definitely larger than
the value predicted by Caracciolo et al 1.5420(7).
\item There is no obvious reason to suspect that the continuum limit in
the two models is different. 
\item The dashed lines represent linear extrapolations in $1/(\ln L+0.7)$
and they intersect at $\xi(2L)/\xi(L)=1.584$. The extrapolations used
and the value they produce should  be regarded only as an illustration, 
since we have no way of knowing whether our $L$ values are truly 
asymptotic and the extrapolation ansatz is correct.
\end{itemize}
 
\begin{figure}[htb]
\centerline{\epsfxsize=10.0cm\epsfbox{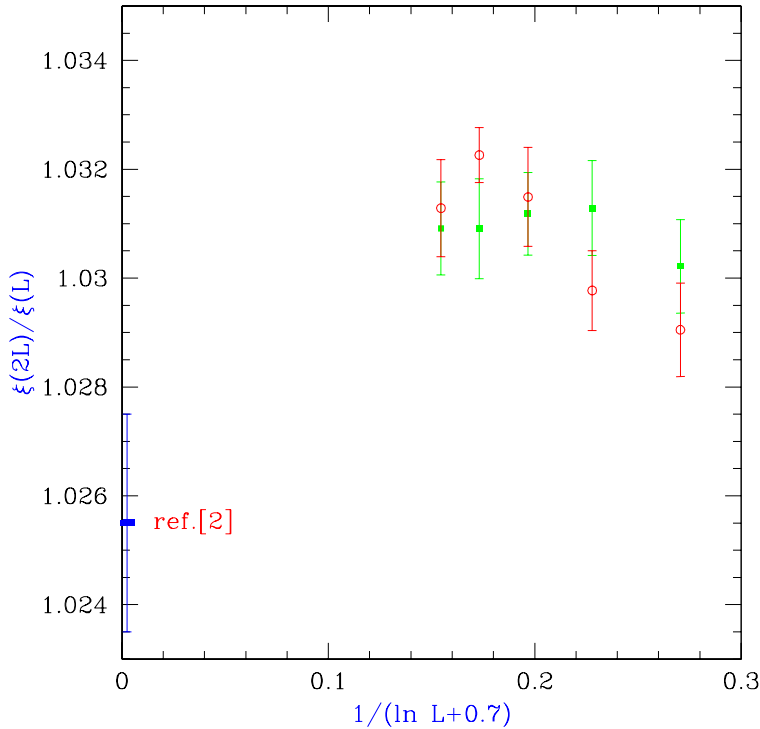}}
\caption{Scaling function $\xi(2L)/\xi(L)$ versus
$1/(\ln L+0.7)$ for $O(3)$ (full squares) and
icosahedron (open circles)        at $\xi(L)/L=0.25$}
\label{step25}
\end{figure}

For a better test of the approach to the continuum, we investigated a smaller
value of $x$, namely $x=0.25$, where one would expect the continuum limit
to be reached at smaller $L$ values. The data are shown in Fig.
\ref{step25}.

\begin{figure}[htb]
\centerline{\epsfxsize=10.0cm\epsfbox{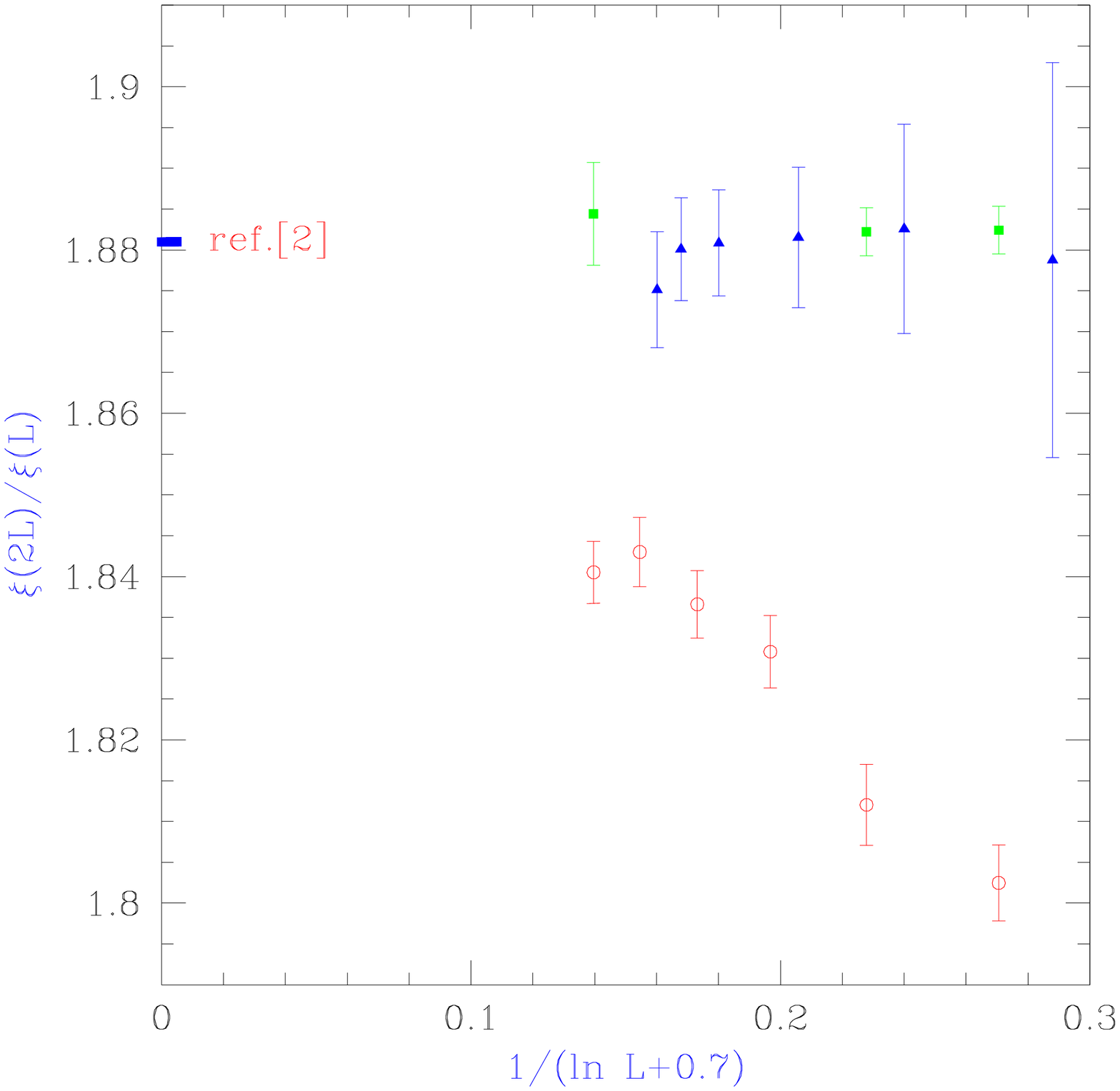}}
\caption{Scaling function $\xi(2L)/\xi(L)$ versus
$1/(\ln L+0.7)$ for $O(3)$ (full squares:our data, full triangles:
ref.\cite{car}) and
icosahedron (open circles) at $\xi(L)/L=0.75$}
\label{step75}
\end{figure}

They suggest the following facts:
\begin{itemize}
\item Indeed the continuum limit seems to be reached at lower $L$ values,
perhaps as low as $L=320$ corresponding to a correlation length $\xi=80$.
\item The data are consistent with the two models sharing a common continuum
limit value for the step scaling function.
\ item This value seems to be around 1.031, higher than the value
quoted by Caracciolo et al (1.0255(20)).
\end{itemize}

In Fig.\ref{step75} we present the same results as in in Fig.\ref{step5} but
at $x=0.75$. To go to this increased value of $x$, we had to increase
$\beta$. Consequently it is to be expected that at this $x$ value, asymptopia
will set in at a much larger value of $L$. This must be so especially for the
$O(3)$ model, which will be described by PT up to much larger values of 
$L$ (as we emphsized above, in the true continuum limit, the system 
cannot be in a PT regime, since in the continuum limit the spins 
decorrelate over distances proportional to $L$).
Two things can nevertheless be learned from this figure:
\begin{itemize}
\item There is no reason to rule out that the continuum value
is the same in the two models.
\item The continuum limit may very well be again higher than the 
prediction of
Caracciolo et al (1.8810(3)), which most likely comes from a transient,
PT dominated regime; with present day's computers one cannot study
reliably the larger $L$ values needed to settle this issue.
\end{itemize}

Our conclusion is that there is no evidence for the crossover phenomenon
in the discrete nonabelian symmetry enhancement suggested by Caracciolo et al.
While it follows from the accepted PT calculations, their scenario seems
bizarre from the point of view of the discrete spin model. Indeed the latter
may simulate a continuous symmetry only if the spins are correlated over a 
sufficiently large portion of the lattice. Fluctuations alone are not 
sufficient to enhance the symmetry, as can be seen by the high 
temperature expansion, which clearly is different for the icosahedron and
the $O(3)$ models.
Symmetry enhancement requires the collective effect of many discrete
spins over the typical distance of a correlation length; therefore
one should expect that the discrete spins are capable of approximating
the continuous ones better and better with increasing correlation length.
Thus it would be bizarre if as the correlation length increases, beyond 
some value, correlation functions at distances measured in units of the
correlation length would become less symmetric. However, being based 
only on some numerics, the present work cannot rule out the Caracciolo 
et al scenario.

What the present work does show though is that the scaling curve predicted
by Caracciolo et al in 1995 \cite{car} is not correct and that the lattice
artefacts are larger than they claimed. If one accepts the universality
of the $O(3)$ and icosahedron spin model, then it appears that one could
get a much better approximation of the continuum scaling curve by
studying the icosahedron spin model. For instance the scaling curve produced 
by Caracciolo et al \cite{car} never reached 2, presumably because the best
it could do is reach the value predicted by perturbation theory. Now in the
icosahedron model one can work at $\beta_{crt}$. The results, shown in 
Tab.7 show that $\xi(2L)/\xi(L)$ does actually reach 2 (definition of
$\beta_{crt}$) and that this happens at $x\sim 1$ (nontrivial information).
If the scaling curve were known, one could repeat the Kim-Caracciolo et al
procedure and produce some thermodynamic values for $\xi(\beta)$. It should
be emphasized though that with that procedure, the error compounds and
one can lose control very fast. For instance if the scaling curve
predicted by Caracciolo et al is changed by adding to their fit formula
for $\xi(2L)/\xi(L)$ the term $0.02x$  and the starting values were
$L=100$ $\xi=75$ then the predicted thermodynamic value changes from 2000
to 2670. Please notice that the modification we introduced vanishes for 
$x\to 0$ and it amounts to an increase of less than 1\% for any $x$, yet the
change in the predicted value is about $33\%$. Since in their 1995 paper
Caracciolo et al stated themselves that they did observe scaling violations 
of about $1.5\%$ (and the data shown in Fig.1 suggest that that was a gross
underestimate), we believe that their claim of having verified AF at
correlation lenghts $10^5$ at the $4\%$ level is unjustified.
\vglue5mm \noindent
{\it Note Added in Proof:}

After the completion of this paper, S. Caracciolo drew our attention to
some recent work done by himself with A. Montanari and A. Pelissetto
concerning the issue of nonabelian discrete symmetry enhancement 
\cite{latt,jhep}. In this work Caracciolo et al claim to have
shown that in fact the continuum limit of the model perturbed by a
term enjoying only icosadral symmetry is different from that of the $O(3)$
model; but in contrast to their earlier estimate that the difference
should show up at correlation length of about 200 they now say that
they have to go to correlation length of about $10^5$.

Unfortunately the authors base their conclusion {\it only} on data
taken on relatively small lattices (while their estimate of the 
thermodynamic correlation length is $\approx 10^5$, their largest
lattice for which they determine the step scaling function only is
$150\times 150$. As we already stated both in previous papers \cite{comk},
\cite{coms}, \cite{jsp} and repeated throughout the present paper, such
data cannot be considered as representing the true continuum limit since
they place the system in a perturbative regime, known rigorously not to
occur in the true continuum limit. It is also not clear if the perturbed
system with the icosahedral symmetry is not in its magnetized phase at
the values of the paramters chosen (the authors do not provide any 
infromation on this).

The data reported in the present paper, at $x=0.25$ and $x=0.5$, do not
suffer from this limitation and are consistent with discrete symmetry
enhancement. In support of their claim, Caracciolo et al state that their
data indicate that the difference between the model enjoying a discrete
symmetry and $O(3)$ is increasing with the size of the lattice $L$.
We have two comments to this observation:
\begin{itemize}
\item Even though the continuum limit of the two models may be the same, there
is no guarantee that this limit {\it must} be reached in a monotonic fashion,
hence the observed increased descrepancy with increased $L$ could be a transient
phenomenon.
\item In some cases (Fig.2 of \cite{latt}, Figs. 4 and 6 of \cite{jhep})
the data at their largest $L$ value $L=150$ seem to lie {\it below} the 
data at $L=90$, contrary to their claim.
\end{itemize}

We thank S. Caracciolo and A. Sokal for correspondence and discussions.
A.P. wishes to thank the Werner Heisenberg Institut, where this study was
conducted, for its hospitality.

\newpage
\begin{center}
{\bf Tab.1:}\\
{\it $\xi(L)$ and $\xi(2L)$ for the $O(3)$ model at 
$\xi(L)/L\approx .25$}\\
\begin{tabular}[t]{r||r|r|r|r|r}
$L$     &  20   & 40    & 80     & 160   &   320\\
\hline
$\beta$ & 1.332 & 1.488 & 1.6135 & 1.728 & 1.838 \\
\hline
$\xi(L)$ &5.011(3) &10.070(7) & 19.919(11)&39.881(25)&79.988(48)\\ 
$\xi(2L)$&5.164(3) &10.395(5) &20.529(10) &41.097(26)&82.458(47)\\  
\end{tabular}\\

\vskip3mm
{\bf Tab.2:}\\
{\it $\xi(L)$ and $\xi(2L)$ for the $O(3)$ model at $\xi(L)/L\approx .5 
$}\\
\begin{tabular}[t]{r||r|r|r|r|r|r}
$L$     &  20   & 40    & 80     & 160   &   320  &  640\\
\hline
$\beta$ & 1.595 & 1.7143& 1.825  & 1.935 & 2.047  & 2.16 \\
\hline
$\xi(L)$ &10.007(9)&19.999(17)& 40.002(28)&79.884(52)&159.73(13)
&319.29(37)\\
$\xi(2L)$&15.205(13)&30.629(27)&61.451(41)&123.079(91)&246.99(26)
&493.94(68)\\
\end{tabular}\\

\vskip3mm
{\bf Tab.3:}\\
{\it $\xi(L)$ and $\xi(2L)$ for the $O(3)$ model at $\xi(L)/L\approx .75$}\\
\begin{tabular}[t]{r||r|r|r}
$L$     &  20   & 40    & 640\\
\hline
$\beta$ & 2.092 & 2.18  & 2.63\\
\hline
$\xi(L)$ &15.010(16)&29.607(30)&470.07(86)\\
$\xi(2L)$&28.260(32)&55.608(66)&882.8(2.5)\\
\end{tabular}\\

\vskip3mm
{\bf Tab.4:}\\
{\it $\xi(L)$ and $\xi(2L)$ for the $icosahedron$ model at 
$\xi(L)/L\approx .25$}\\
\begin{tabular}[t]{r||r|r|r|r|r}
$L$     &  20   & 40    & 80     & 160   &   320\\
\hline
$\beta$ & 1.319 & 1.457 & 1.556  & 1.6295& 1.6836\\
\hline
$\xi(L)$ &5.016(3) &10.043(6) & 19.960(14)&39.955(15)&80.278(57)\\
$\xi(2L)$&5.164(3) &10.348(4) &20.583(11) &41.237(13)&82.829(41)\\
\end{tabular}\\

\vskip3mm
{\bf Tab.5:}\\
{\it $\xi(L)$ and $\xi(2L)$ for the $icosahedron$ model at 
$\xi(L)/L\approx .5 $}\\
\begin{tabular}[t]{r||r|r|r|r|r|r}
$L$     &  20   & 40    & 80     & 160   &   320  &  640\\
\hline
$\beta$ & 1.5382& 1.618 & 1.6767 & 1.718 &   1.7469 &  1.7664\\
\hline
$\xi(L)$ &9.991(9)  & 19.996(17) &40.057(25) & 80.131(74)&160.04(20)
& 318.62(32)\\
$\xi(2L)$&14.955(14)& 30.141(20) & 61.076(42)& 122.58(13) &246.14(35)
& 487.17(70)\\
\end{tabular}\\

\vskip3mm
{\bf Tab.6:}\\
{\it $\xi(L)$ and $\xi(2L)$ for the $icosahedron$ model at 
$\xi(L)/L\approx .75$}\\
\begin{tabular}[t]{r||r|r|r|r|r|r}
$L$     &  20   & 40    & 80     & 160   &   320  &  640\\
\hline
$\beta$ & 1.733 & 1.758 & 1.774  & 1.788 & 1.794 & 1.797 \\
\hline
$\xi(L)$ & 15.114(31) & 29.936(68) & 58.99(11)&121.20(21)&242.68(39)
&479.40(5) \\
$\xi(2L)$& 27.279(43) & 54.227(083)&107.69(17)&222.97(33)&448.06(73)
&882.2(1.6)\\
\end{tabular}\\

\vskip3mm
{\bf Tab.7a:}\\
{\it $\xi(L)$ for the  icosahedron  model at $\beta=1.802$
}
\begin{tabular}[t]{r||r|r|r|r|r}
$L$     &  20   & 40    & 80     & 160   &   320  \\
\hline
$\xi(L)$ & 19.946(49) & 39.50(11) & 78.11(21)&155.77(38)&305.20(51)
\\
\end{tabular}\\
\vskip3mm
{\bf Tab.7b:}\\
{\it $\xi(L)$ for the  icosahedron  model at
$\beta=1.803$}
\begin{tabular}[t]{r||r|r}
$L$     &   320   & 640  \\
\hline
$\xi(L)$ & 
318.23(76)&642.2(1.4)\\
\end{tabular}\\
\end{center}

\end{document}